\documentclass[aps,twocolumn,amsmath]{revtex4}
\usepackage{amssymb}
\usepackage{graphicx}
\usepackage{array}
\usepackage{hhline}
\usepackage{longtable}

\renewcommand{\thetable}{\Roman{table}} \thetable

\begin{document}

\title{Phase Diagrams and Crossover in

Spatially Anisotropic d=3 Ising, XY Magnetic and Percolation
Systems:

Exact Renormalization-Group Solutions of Hierarchical Models}
\author{Aykut Erba\c{s},$^1$ Asl\i  ~Tuncer,$^1$ Burcu Y\"ucesoy,$^1$ and A. Nihat Berker$^{1-3}$}
\affiliation{$^1$Department of Physics, Istanbul Technical
University, Maslak 34469, Istanbul, Turkey,}
\affiliation{$^2$Department of Physics, Massachusetts Institute of
Technology, Cambridge, Massachusetts 02139, U.S.A.,}
\affiliation{$^3$Feza G\"ursey Research Institute, T\"UBITAK -
Bosphorus University, \c{C}engelk\"oy 81220, Istanbul, Turkey}
\begin{abstract}
Hierarchical lattices that constitute spatially anisotropic systems
are introduced.  These lattices provide exact solutions for
hierarchical models and, simultaneously, approximate solutions for
uniaxially or fully anisotropic $d=3$ physical models.  The global
phase diagrams, with $d=2$ and $d=1$ to $d=3$ crossovers, are
obtained for Ising, XY magnetic models and percolation systems,
including crossovers from algebraic order to true long-range order.

\end{abstract}
\maketitle
\def\s{\rule{0in}{0.28in}}

\section{Introduction}

\setlength{\LTcapwidth}{\columnwidth}

Spatially anisotropic systems greatly enrich our experience of
collective phenomena, as exemplified by high-$T_c$ superconducting
materials, in which the couplings along one direction are much
weaker than those in the perpendicular plane.  Anisotropic systems
are also intriguing from a conceptual point of view, since vastly
different critical phenomena are known to happen in different
spatial dimensions, whereas, between $d$-dimensional systems stacked
along a new direction, even the weakest coupling, while not
affecting the critical temperature, induces $(d+1)$-dimensional
critical behavior. Calculational results that yield the global phase
diagram of anisotropic systems and thus provide a unified connected
picture of the various anisotropic and isotropic behaviors at
different subdimensions and at the full dimension have been rare and
mostly confined to $d=2$.  In the present study, we obtain global
phase diagrams for a variety of anisotropic $d=3$ systems: Ising
magnetic, XY magnetic, and percolation systems.  Anisotropy along
one direction (uniaxial) and full anisotropy, in which the couplings
along each direction are different, are studied, yielding global
phase diagrams.  We use hierarchical models, which yield exact
renormalization-group solutions \cite{Berker, Kaufman, Kaufman2}.
Thus, the construction of hierarchical lattices that incorporate
correct dimensional reductions is an important step of the study.
The exact solutions of hierarchical models can simultaneously be
considered approximate position-space renormalization-group
solutions of models on naturally occurring lattices \cite{Berker}.
The method developed in this study will be employed to extend, from
isotropic to anisotropic systems, the renormalization-group
solutions of the tJ and Hubbard models of electronic conduction
\cite{Falicov, Hinczewski, Hinczewski2}.

\section{Anisotropic Hierarchical Lattices}

Hierarchical lattices are constructed by repeatedly self-imbedding a
graph.  These provide exactly solvable models, with which complex
problems can be studied and understood.  For example, frustrated
\cite{McKay}, spin-glass \cite{Migliorini}, random-bond
\cite{Andelman} and random-field \cite{Falicov2}, Schr\"odinger
equation \cite{Domany}, lattice-vibration \cite{Langlois}, dynamic
scaling \cite{Stinchcombe}, aperiodic magnet \cite{Haddad}, complex
phase diagram \cite{Le}, and directed-path \cite{daSilveira}
systems, etc., have been solved on hierarchical lattices.

\begin{figure}
\centering

\includegraphics*[scale=0.5]{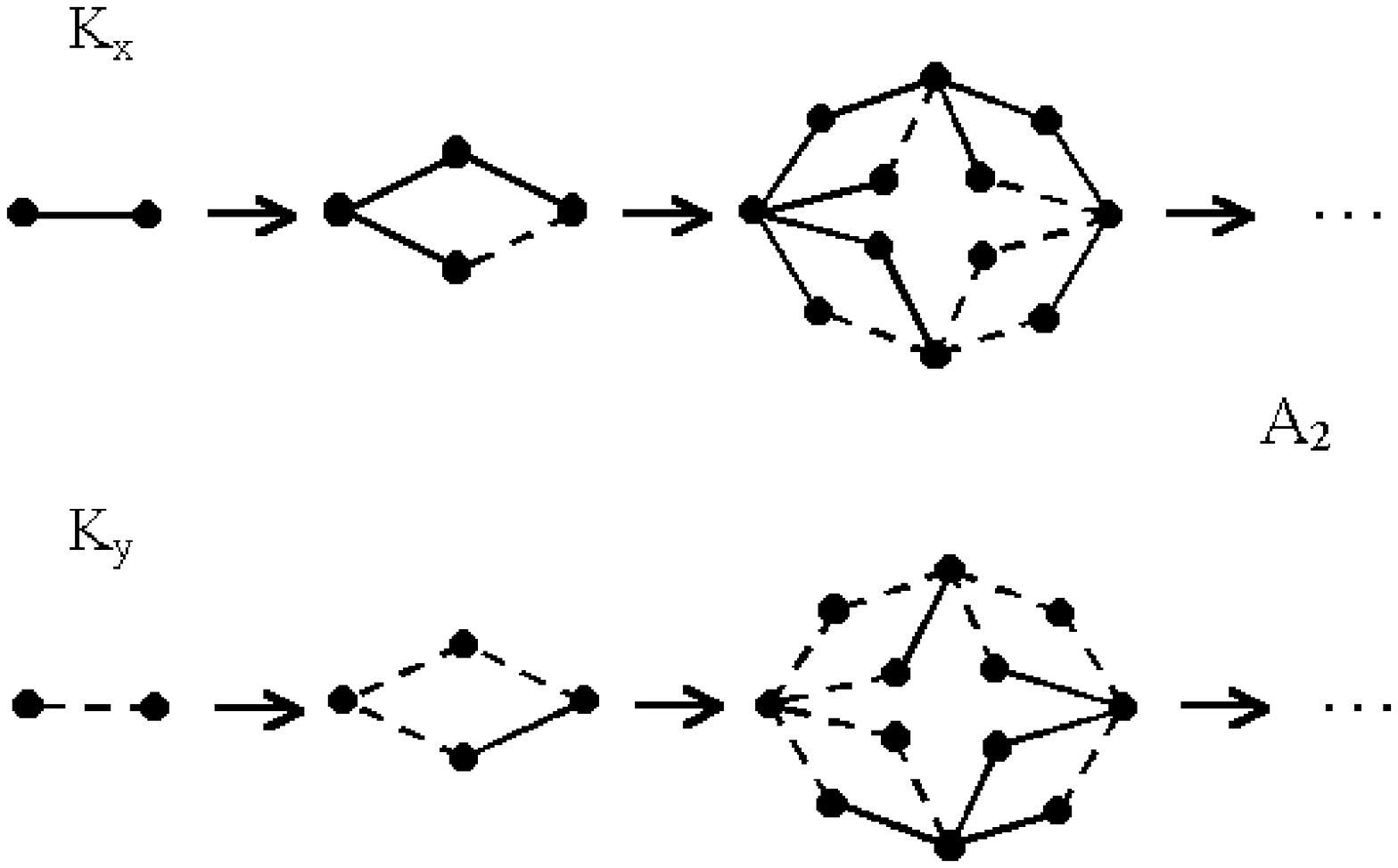}

\caption{The parallel, mutual imbeddings of these two graphs provide
an anisotropic $d=2$ hierarchical lattice.  If either of the
couplings $(K_x,K_y)$ is set to zero, the remaining coupling
constitutes a one-dimensional lattice.  When the couplings are of
equal strength, $K_x = K_y$, the two directions, represented by the
two imbedding sequences, are equivalent and the lattice is isotropic
$d=2$. This lattice will be referred to as $A_2$.}
\end{figure}

In this study, we construct anisotropic hierarchical lattices by the
parallel, mutual imbedding of several graphs.  In each imbedding
step, $b$ and $b^d$ respectively are the length and volume rescaling
factors.  We illustrate the method by the simplest case of the
anisotropic $d=2$ lattice, before moving on to the uniaxially or
fully anisotropic $d=3$ lattices.  The parallel, mutual imbeddings
of the two graphs shown in Fig.1 provide an anisotropic $d=2$
hierarchical lattice. If either of the couplings ($K_x,K_y$) is set
to zero, the remaining coupling constitutes a one-dimensional
lattice.  When the couplings are of equal strength, $K_x = K_y$, the
two directions, represented by the two imbedding sequences, are
equivalent and the lattice is isotropic $d=2$. This lattice will be
referred to as $A_2$.

It is thus seen that generally our requirements in the construction
of anisotropic hierarchical lattices are (1) the proper reduction to
the lower dimension when one (or more, see below) of the couplings
is set to zero and (2) the restitution of an isotropic lattice when
the couplings are of equal strength.

\begin{figure}
\centering

\includegraphics*[scale=0.5]{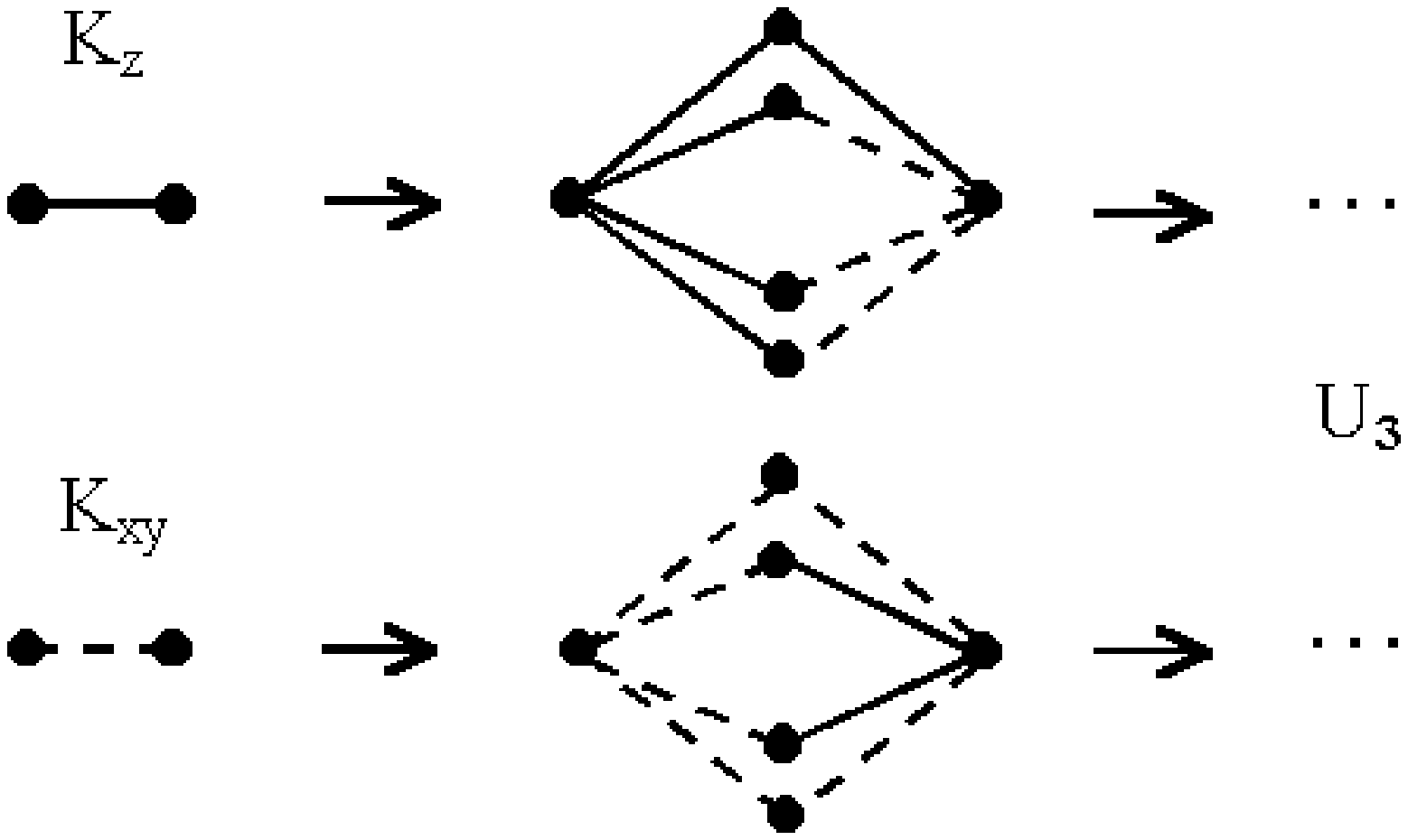}

\caption{The parallel, mutual imbeddings of these two graphs provide
a uniaxially anisotropic $d=3$ hierarchical lattice.  If the
coupling $K_z$ is set to zero, the coupling $K_{xy}$ constitutes an
isotropic two-dimensional lattice.  If the coupling $K_{xy}$ is set
to zero, the coupling $K_z$ constitutes a one-dimensional lattice.
When the couplings are of equal strength, $K_{xy} = K_z$, the $z$
direction, represented by the first imbedding sequence, and the
$x,y$ directions, represented by the second imbedding sequence, are
all equivalent and the lattice is isotropic $d=3$.  This lattice
will be referred to as $U_3$.}
\end{figure}

\begin{figure}
\centering

\includegraphics*[scale=0.59]{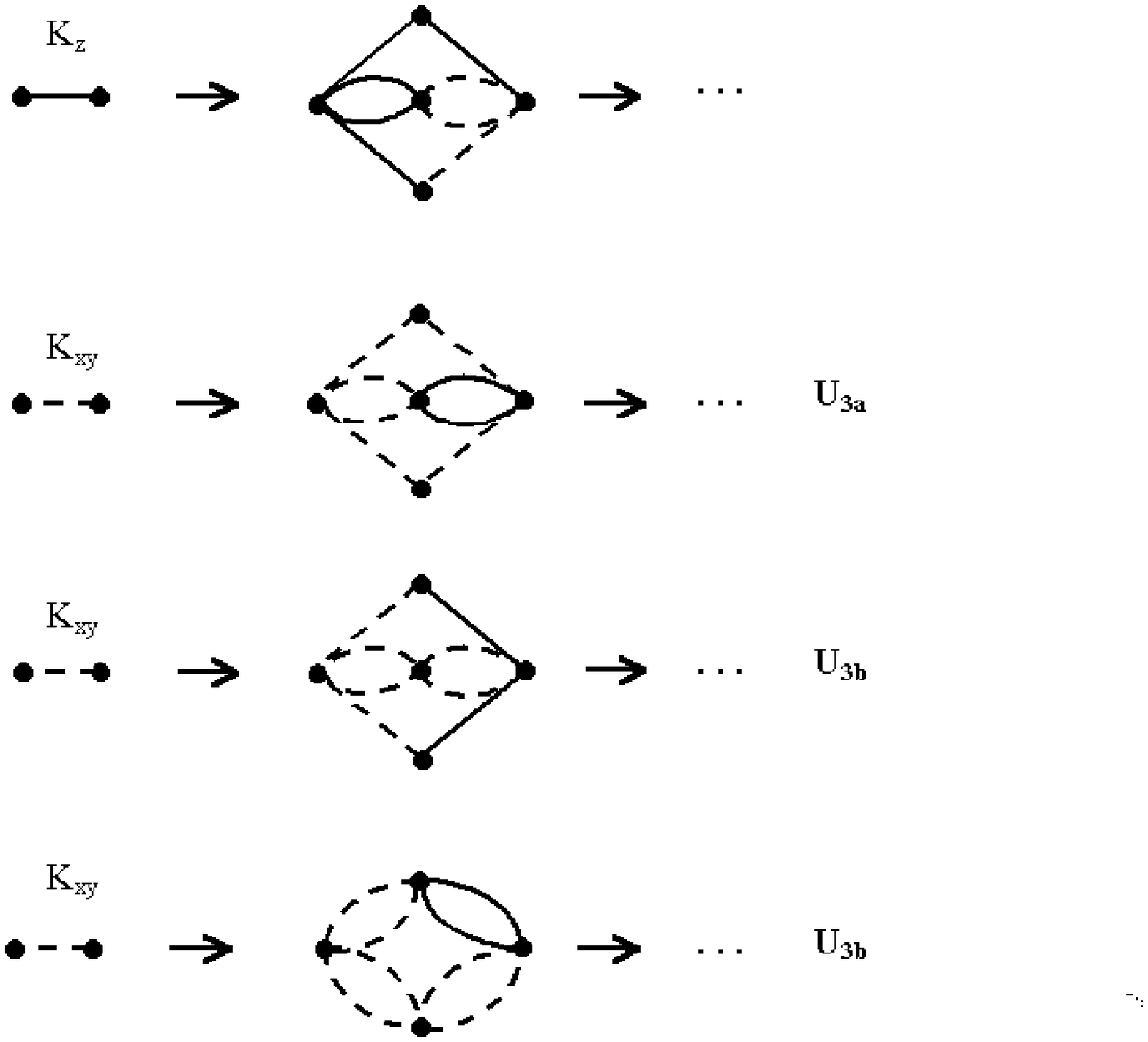}

\caption{The parallel, mutual imbeddings of the top graph with
either one of the following graphs provide uniaxially anisotropic
$d=3$ hierarchical lattices.  If the last graph is used, isotropy is
not restored when $K_{xy} = K_z$.  These lattices, differentiated by
the choice of the second imbedding graph, will be respectively
referred to as $U_{3a}, U_{3b}, U_{3c}$.}
\end{figure}

The parallel, mutual imbeddings of the two graphs shown in Fig.2
provide a uniaxially anisotropic $d=3$ hierarchical lattice.  If the
coupling $K_z$ is set to zero, the coupling $K_{xy}$ constitutes an
isotropic two-dimensional lattice.  If the coupling $K_{xy}$ is set
to zero, the coupling $K_z$ constitutes a one-dimensional lattice.
When the couplings are of equal strength, $K_{xy} = K_z$, the $z$
direction, represented by the first imbedding sequence, and the
$x,y$ directions, represented by the second imbedding sequence, are
all equivalent and the lattice is isotropic $d=3$.  This lattice
will be referred to as $U_3$.  In Fig.3, the parallel, mutual
imbeddings of the top graph with either one of the following graphs
also provide uniaxially anisotropic $d=3$ hierarchical lattices.  If
the last graph is used, isotropy is not restored when $K_{xy} =
K_z$; this lattice is nevertheless included, for comparison, in our
study. These lattices, differentiated by the choice of the second
imbedding graph, will be respectively referred to as $U_{3a},
U_{3b}, U_{3c}$.

\begin{figure}
\centering

\includegraphics*[scale=0.47]{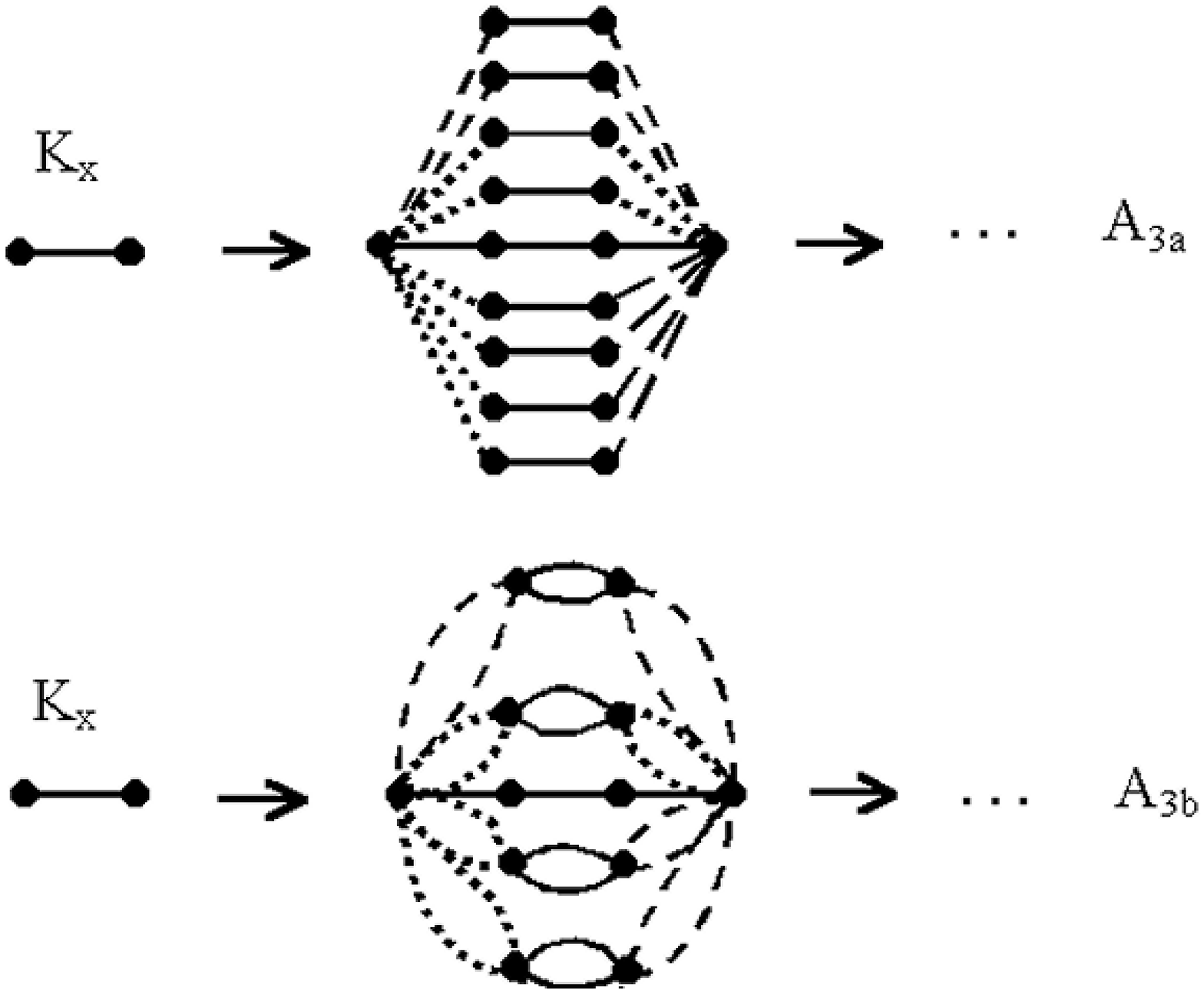}

\caption{Each imbedding shown in this figure, in parallel with the
two imbeddings obtained by permuting $K_x$ (full line), $K_y$
(dashed), and $K_z$(dotted), yields a fully anisotropic $d=3$
hierarchical lattice. If any one the couplings $K_u$ is set to zero,
the remaining two couplings constitute an anisotropic
two-dimensional lattice.  If any two of the couplings are set to
zero, the remaining coupling constitutes a one-dimensional lattice.
When the couplings are of equal strength, $K_x = K_y = K_z$, the
three directions, represented by the three mutual imbedding
sequences, are equivalent and the lattice is isotropic $d=3$. These
lattices will be respectively referred to as $A_{3a}$ and $A_{3b}$.}
\end{figure}

A fully anisotropic $d=3$ hierarchical lattice is provided in Fig.4
by each shown imbedding in parallel with the two imbeddings obtained
by permuting $K_x$ (full line), $K_y$ (dashed), and $K_z$ (dotted).
If any one the couplings $K_u$ is set to zero, the remaining two
couplings constitute an anisotropic two-dimensional lattice.  If any
two of the couplings are set to zero, the remaining coupling
constitutes a one-dimensional lattice. When the couplings are of
equal strength, $K_x = K_y = K_z$, the three directions, represented
by the three imbedding sequences, are equivalent and the lattice is
isotropic $d=3$.  These lattices will be referred to as $A_{3a}$ and
$A_{3b}$.

The anisotropic systems that we study are located on the anisotropic
lattices constructed above.  These hierarchical models admit exact
renormalization-group solutions, with recursion relations obtained
by decimations in direction opposite to their construction
direction.  The exact solutions of hierarchical models can
simultaneously be considered approximate position-space
renormalization-group solutions of models on naturally occurring
lattices.  In fact, the recursion relations obtained for the models
below correspond to Migdal-Kadanoff \cite{Migdal, Kadanoff}
approximate recursion relations, which are hereby generalized to
anisotropic systems.

\newpage

\section{Anisotropic Ising Magnets}

The Ising model is defined by the Hamiltonian
\begin{align}
-\beta H=&\sum_{u}K_u\sum_{\langle ij\rangle_u}s_i s_j,\label{eq:1}
\end{align}
where, at each lattice site $i$, $s_i = \pm 1$, and $<ij>_u$ denotes
summation over bonds of type $u$.  The various decimations in the
models are composed of two elementary steps, $K = K_u + K_v$ for
bonds in parallel and $K = \tanh^{-1}(\tanh K_u + \tanh K_v)$ for
bonds in series, where $K$ is the effective coupling of the combined
bonds.

The phase boundaries for the Ising model on the $d=2$ anisotropic
hierarchical lattices $A_2, A_{3a},$ and $A_{3b}$ (setting $K_z=0$
in the latter two) are given in Fig.5, along with the exact result
for the anisotropic square lattice \cite{Onsager}.  The
renormalization-group flows are indicated on the phase boundary of
the hierarchical models.  The fixed point occurs at isotropy, $K_x =
K_y$, to which the $d=2$ anisotropic critical points flow, thereby
sharing the same critical exponents.

The phase boundaries for the Ising model on the $d=3$ uniaxially
anisotropic hierarchical lattices $U_3, U_{3a}, U_{3b}, U_{3c},
A_{3a}$, and $A_{3b}$ (setting $K_x = K_y$) are given in Fig.6. The
exact phase transition points for the square \cite{Onsager} and
cubic \cite{Ferrenberg} lattices are also shown. For each model, the
phase transitions at $d=1$ (at infinite coupling) and $d=2$ cross
over to $d=3$ criticality, which is thus universal for all $d=3$
anisotropic and the $d=3$ isotropic cases.

The phase boundary surface for the Ising model on the $d=3$ fully
anisotropic hierarchical lattice $A_{3b}$ is given in Fig.7.  The
dashed lines on the planes are the exact $d=2$ solutions for the
square lattice \cite{Onsager}.  Again, all points on the critical
surface of the $d=3$ fully anisotropic model flow onto the fixed
point located at isotropy, thereby sharing its critical exponents.
The critical exponents found for this model are $y_T=0.69,
~y_H=1.68$ for $d=2$ (for the square lattice $y_T=1,
~y_H=1.875$~\cite{Onsager}) and $y_T=0.92, ~y_H=2.20$ for $d=3$ (for
the cubic lattice $y_T=1.59, ~y_H=2.50$~\cite{Moore,Gaunt}).

\begin{figure}
\centering
\includegraphics*[scale=0.61]{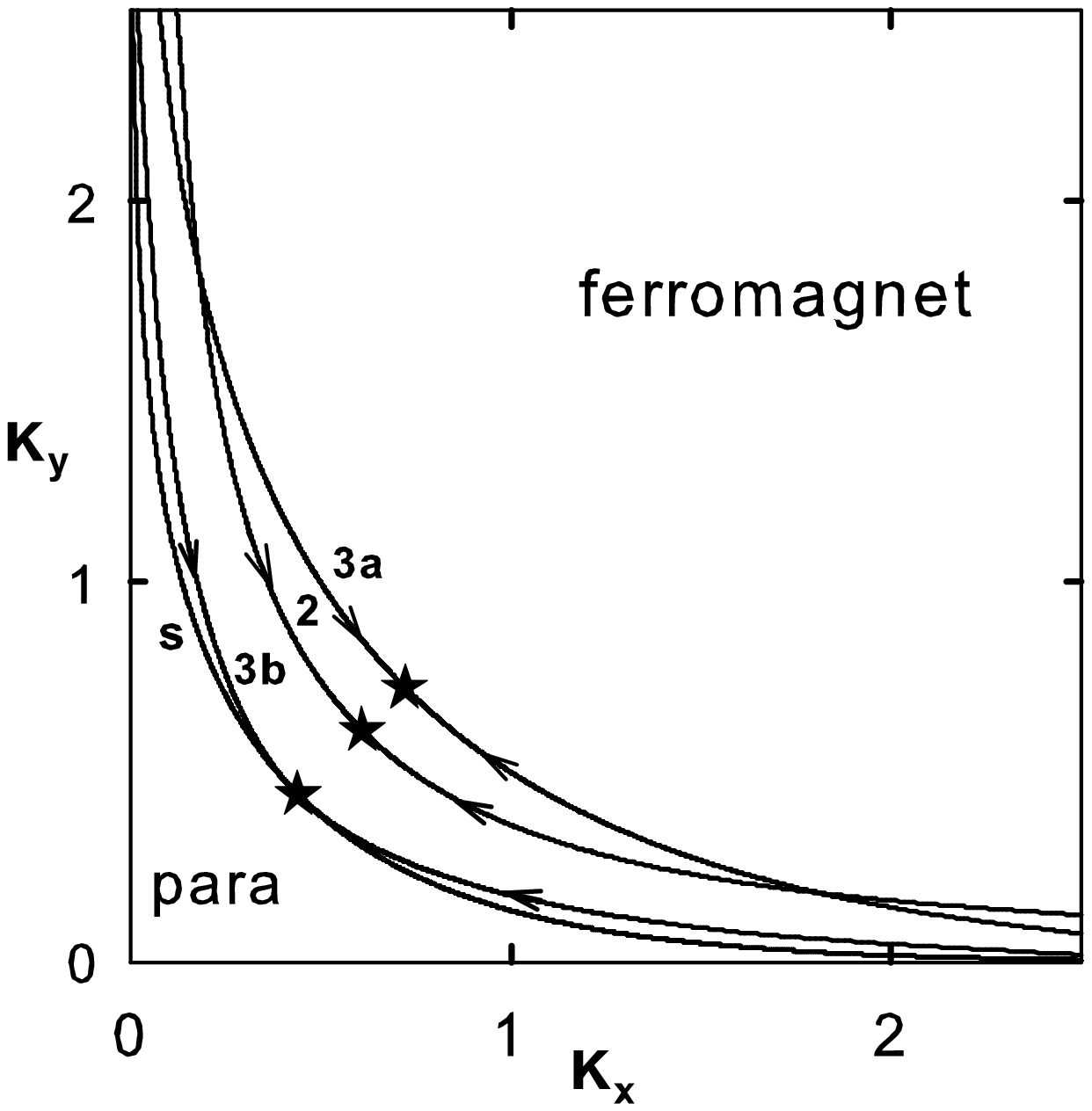}

\caption{The phase boundaries for the Ising model on the $d=2$
anisotropic hierarchical lattices $A_2, A_{3a},$ and $A_{3b}$
(setting $K_z=0$) and the exact result \cite{Onsager} for the
anisotropic square lattice ($s$). The arrows and stars respectively
indicate the renormalization-group flows and fixed points.}
\end{figure}

\begin{figure}
\centering

\includegraphics*[scale=0.99]{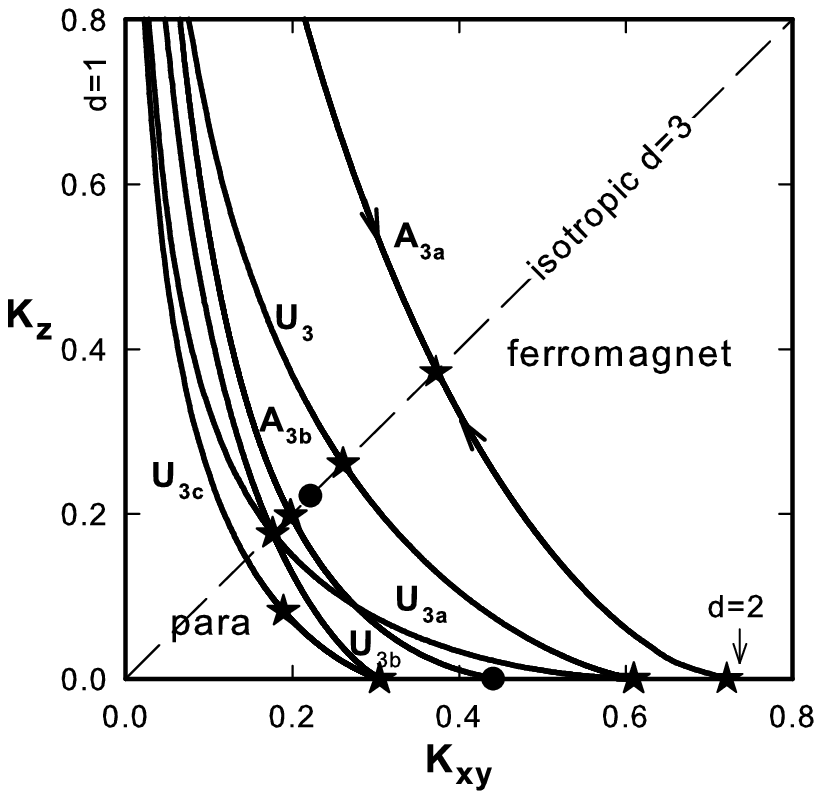}

\caption{Phase boundaries for the Ising model on the $d=3$
uniaxially anisotropic hierarchical lattices $U_3, U_{3a}, U_{3b},
U_{3c}, A_{3a},$ and $A_{3b}$ (setting $K_x = K_y$).  The exact
phase transition points for the square \cite{Onsager} and cubic
\cite{Ferrenberg} lattices are shown by the black circles. For each
model, the phase transitions at $d=1$ (at infinite coupling) and
$d=2$ cross over (as shown for $A_{3a}$) to $d=3$ criticality, which
is thus universal for all $d=3$ anisotropic and the $d=3$ isotropic
cases. The $d=2$ fixed point of $A_{3b}$ is not marked by a star,
since it coincides with the square lattice exact transition point,
marked by the black circle on the horizontal axis.}
\end{figure}

\begin{figure}
\centering

\includegraphics*[scale=0.52]{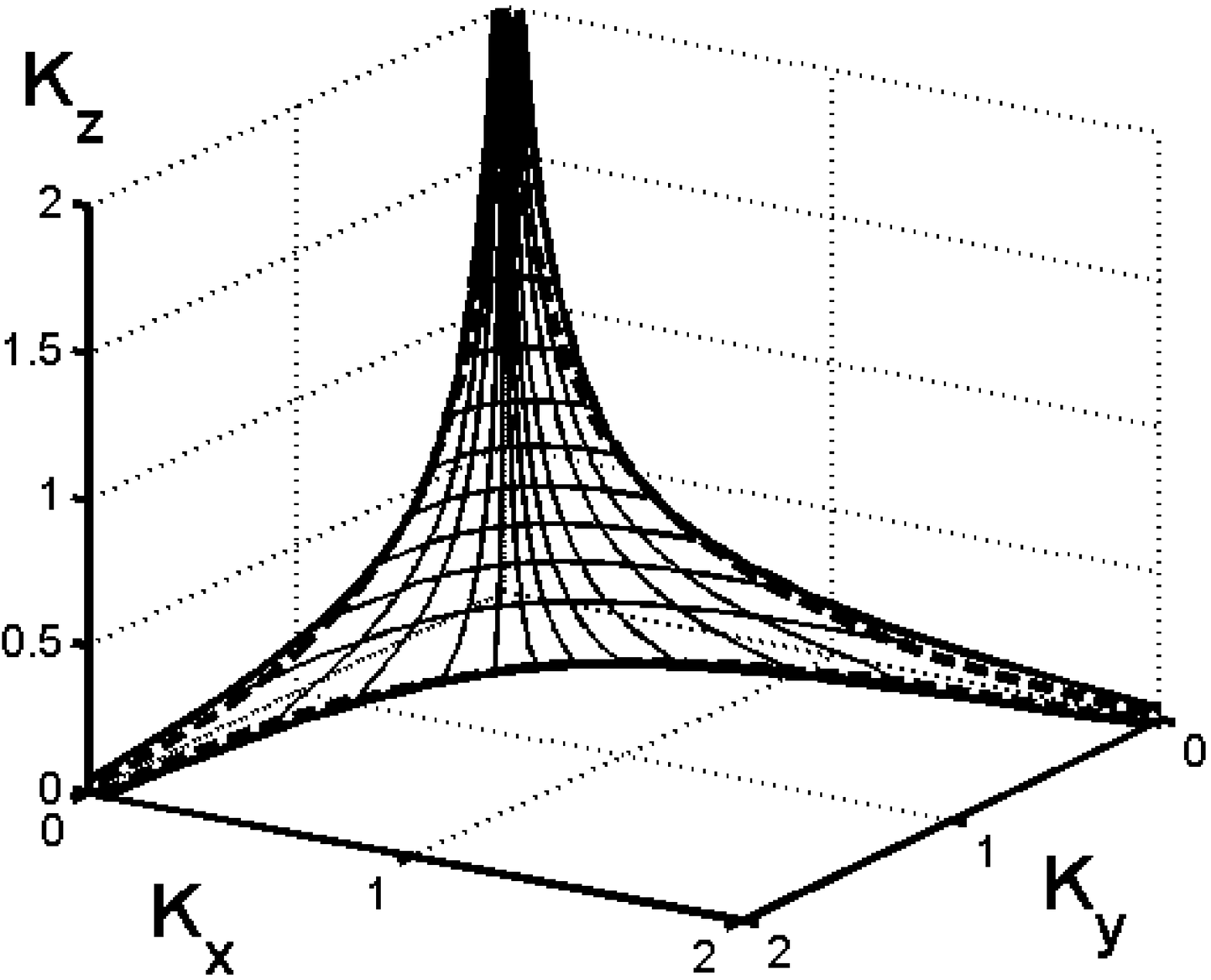}

\caption{Phase boundary surface for the Ising model on the $d=3$
fully anisotropic hierarchical lattice $A_{3b}$.  The dashed lines
on the planes are the exact $d=2$ solutions for the square lattice
\cite{Onsager}.}
\end{figure}

\newpage

\section{Anisotropic XY Magnets}

The XY model is defined by the Hamiltonian
\begin{align}
-\beta H=&\sum_{u}J_u\sum_{\langle ij\rangle_u} \mathbf{s}_i \cdot
\mathbf{s}_j = \sum_{u}J_u\sum_{\langle ij\rangle_u}
\cos(\theta_i-\theta_j),\label{eq:2}
\end{align}
where at each lattice site $i$, $\mathbf{s}_i$ is a unit vector
confined to the $xy$ plane at angle $\theta _i$ to the $x$ axis and
$<ij>_u$ denotes summation over bonds of type $u$.  Under
renormalization-group transformations, the coupling between
nearest-neighbor sites takes the general form of a function
$V_u(\theta _i-\theta _j)$.  The various decimations in the models
are composed of two elementary steps,
\begin{align}
V =& V_u + V_v \quad \text{and}\label{eq:3}\\ V(\theta_i-\theta_k)
=& \ln\int_0^{2\pi} d\theta _j \exp[V_u(\theta_i-\theta_j)+
V_v(\theta_j-\theta_k)],\nonumber
\end{align}
respectively for bonds in parallel and in series, where $V$ is the
effective coupling of the combined bonds.  In terms of Fourier
components,
\begin{align}
f_u(s)=& \int_0^{2\pi} \frac{d\theta}{2\pi} e^{is\theta}
\exp[V_u(\theta)-V_u(0)],\\
\exp[V_u(\theta)-&V_u(0)]  = \sum_{s} e^{-is\theta}
f_u(s),\label{eq:4}\nonumber
\end{align}
Eqs.(3) respectively are
\begin{align}
f(s) =& \sum_{p} f_u(p) f_v(s-p) \quad \text{and}\\ f(s) =&
f_u(s)f_v(s),\label{eq:5}\nonumber
\end{align}
in a form that is more conveniently followed in our calculations.
The phase boundaries for the XY model on the $d=3$ uniaxially
anisotropic hierarchical lattices $U_3, U_{3a}, U_{3b},$ and
$U_{3c}$ are given in Fig.8. The exact phase transition points for
the square \cite{Hasenbusch} and cubic \cite{Ferer} lattices are
also shown.

In $d=2$, namely along the horizontal axis, above a critical
interaction strength marked by the squares on the figure, the
systems exhibit algebraic order \cite{Kosterlitz, Jose, Berker2}:
The starting Hamiltonian [Eq.(2)] flows to a Villain potential
\cite{Villain},
\begin{align}
f_V(s) =& A \exp(-s^2/2J_V),
\end{align}
exhibiting a fixed-line behavior parametrized by $J_V$. This
corresponds to a system without true long-range order, namely with
zero magnetization, but infinite correlation length and algebraic
order in which the correlations decay as a power law. In $d=3$, true
long-range order occurs: points in the ferromagnetic phase
renormalize to a delta function potential; points on the phase
boundaries renormalize to single true fixed potential, shown in
Fig.9, differing from the Villain potential as also seen on the
figure.  The behavior here for $d=2$ is not true fixed-line
behavior.  After tens of thousands of renormalization-group
iterations (corresponding to a scale change factor of 210,000), the
Villain potential decays \cite{Berker2} to a disordered sink with
$V(\theta)_{max}- V(\theta)_{min}<10^{-4}$. The sharp change in the
necessary number of iterations, as seen in Fig.10, indicates the
onset of effective algebraic order.

As seen in Fig.8, for each XY model, the phase transitions at $d=1$
(at infinite coupling) and $d=2$ (onset of algebraic order) cross
over to $d=3$ criticality, which is thus universal for all $d=3$
anisotropic and the $d=3$ isotropic cases.

\begin{figure}
\centering

\includegraphics*[scale=0.95]{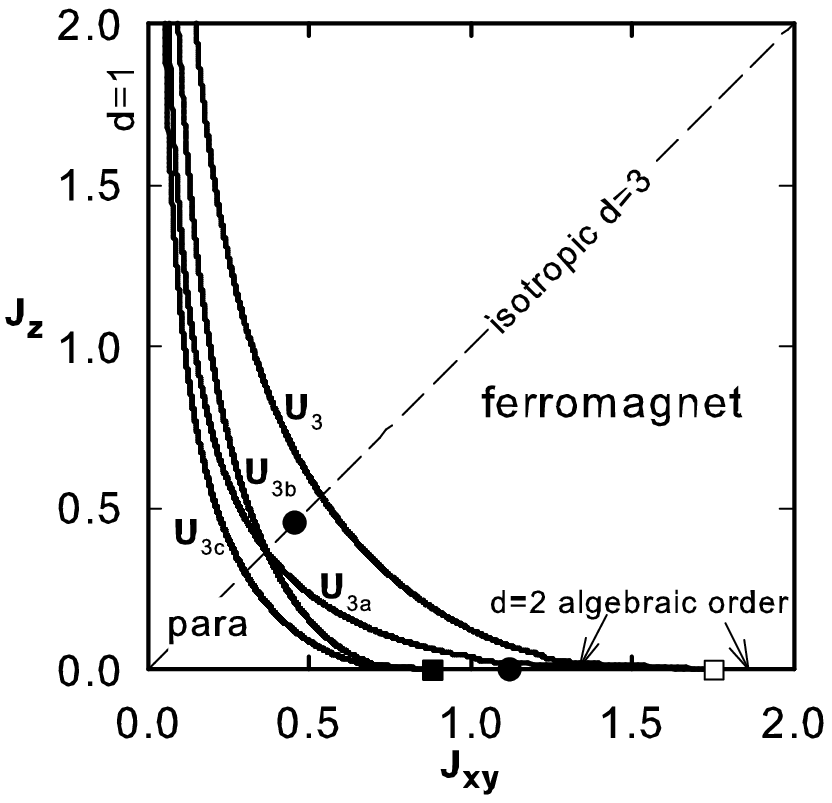}

\caption{Phase boundaries for the XY model on the d=3 uniaxially
anisotropic hierarchical lattices $U_3, U_{3a}, U_{3b}$, and
$U_{3c}$. The exact phase transition points for the square
\cite{Hasenbusch} and cubic \cite{Ferer} lattices are shown by the
black circles. For each model, the phase transitions at $d=1$ (at
infinite coupling) and $d=2$ (onset of algebraic order) cross over
to $d=3$ criticality, which is thus universal for all $d=3$
anisotropic and the $d=3$ isotropic cases.  The onsets of effective
algebraic order in $d=2$ are marked, for models $U_3, U_{3a}$ with
the open square and for models $U_{3b}, U_{3c}$ with the full
square.}
\end{figure}

\begin{figure}
\centering

\includegraphics*[scale=0.9]{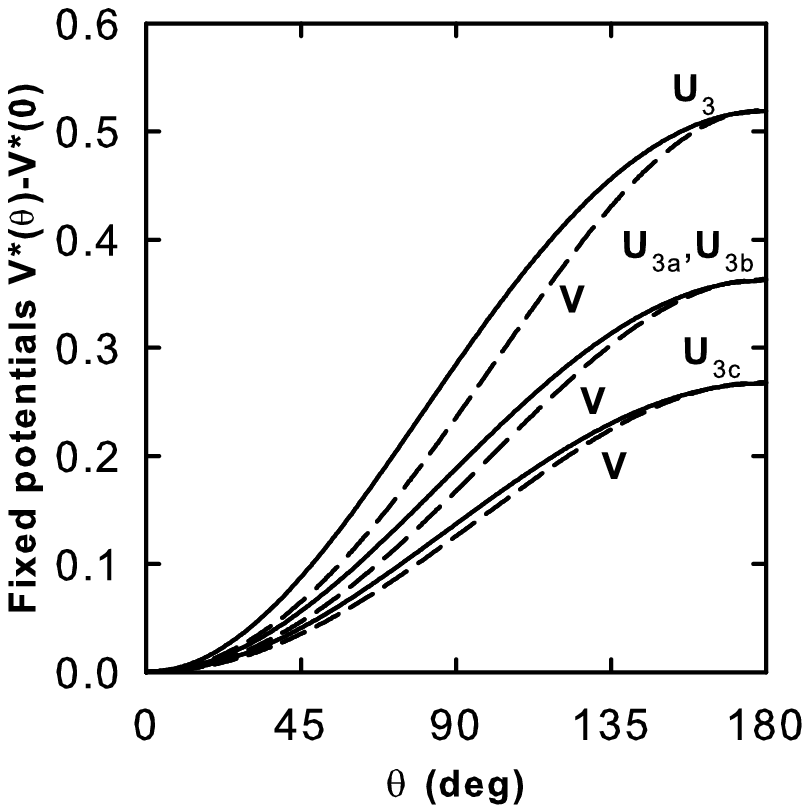}

\caption{Fixed potentials attracting the critical surface of the
$d=3$ XY model.  For comparison, appropriately normalized Villain
potentials are also shown.}
\end{figure}

\begin{figure}
\centering

\includegraphics*[scale=0.9]{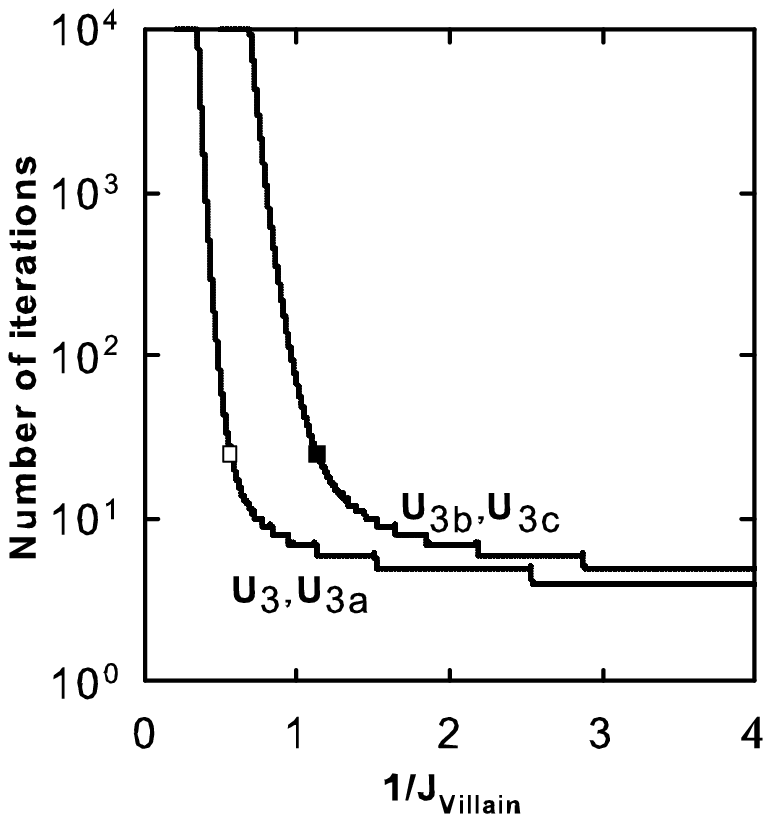}

\caption{Number of renormalization-group iterations necessary, in
the $d=2$ XY model, for the Villain potential parametrized by $J_V$
to decay to a disordered sink with
$V(\theta)_{max}-V(\theta)_{min}<10^{-4}$.  The squares indicate the
onset of effective algebraic order.}
\end{figure}

\section{Anisotropic Percolation}

Anisotropic percolation is defined such that on each connection of
direction $u$, a bond exist with probability $p_u$.  The various
decimations in the models are composed of two elementary steps, $p =
p_up_v + p_u(1-p_v) + p_v(1-p_u)$ for connections in parallel and $p
= p_up_v$ for connections in series, where $p$ is the effective
connectedness probability of the combined connections. The phase
diagram for percolation on the $d=2$ anisotropic hierarchical
lattice ($A_2$) is given in Fig.11.  The percolation fixed point
occurs at isotropy, $p_x = p_y$, to which the $d=2$ anisotropic
percolation onsets flow, thereby sharing the same critical
exponents.  The phase boundaries for percolation on the $d=3$
uniaxially anisotropic hierarchical lattices $U_3, U_{3a}, U_{3b}$,
and $U_{3c}$ are given in Fig.12.  The percolation points for the
cubic and square lattices are also shown \cite{Essam}.  For each
model, percolation onset at $d=1$ (at $p_z =1$) and $d=2$ cross over
to $d=3$ percolation onset, which is thus universal for all $d=3$
anisotropic and the $d=3$ isotropic cases.

\begin{figure}
\centering

\includegraphics*[scale=0.7]{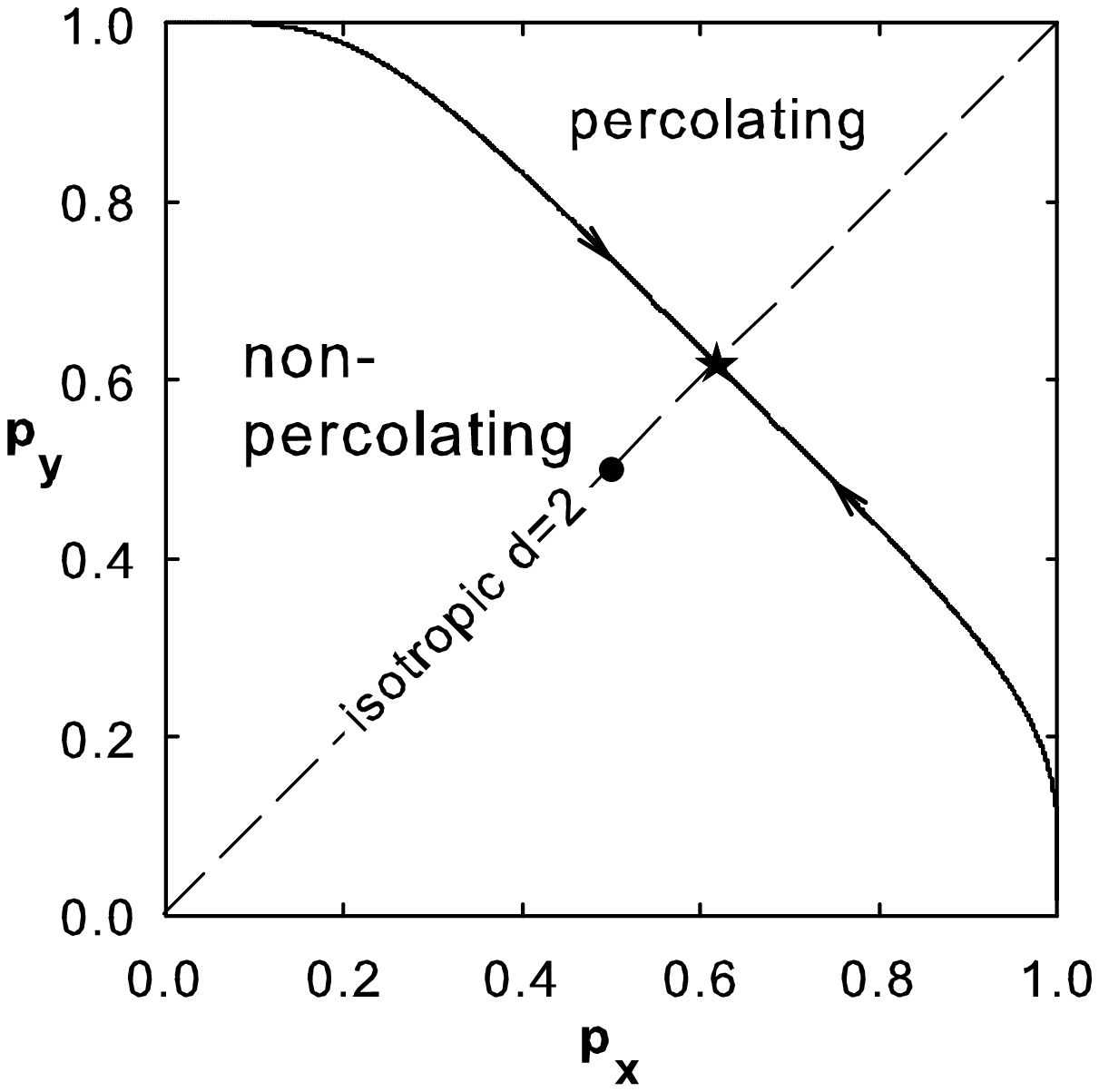}

\caption{The phase boundaries for percolation on the $d=2$
anisotropic hierarchical lattice ($A_2$).  The renormalization-group
flows are indicated on the phase boundary of the hierarchical model.
The onset of percolation for the square lattice \cite{Essam} is
shown by the black circle.}
\end{figure}

\begin{figure}
\centering

\includegraphics*[scale=0.7]{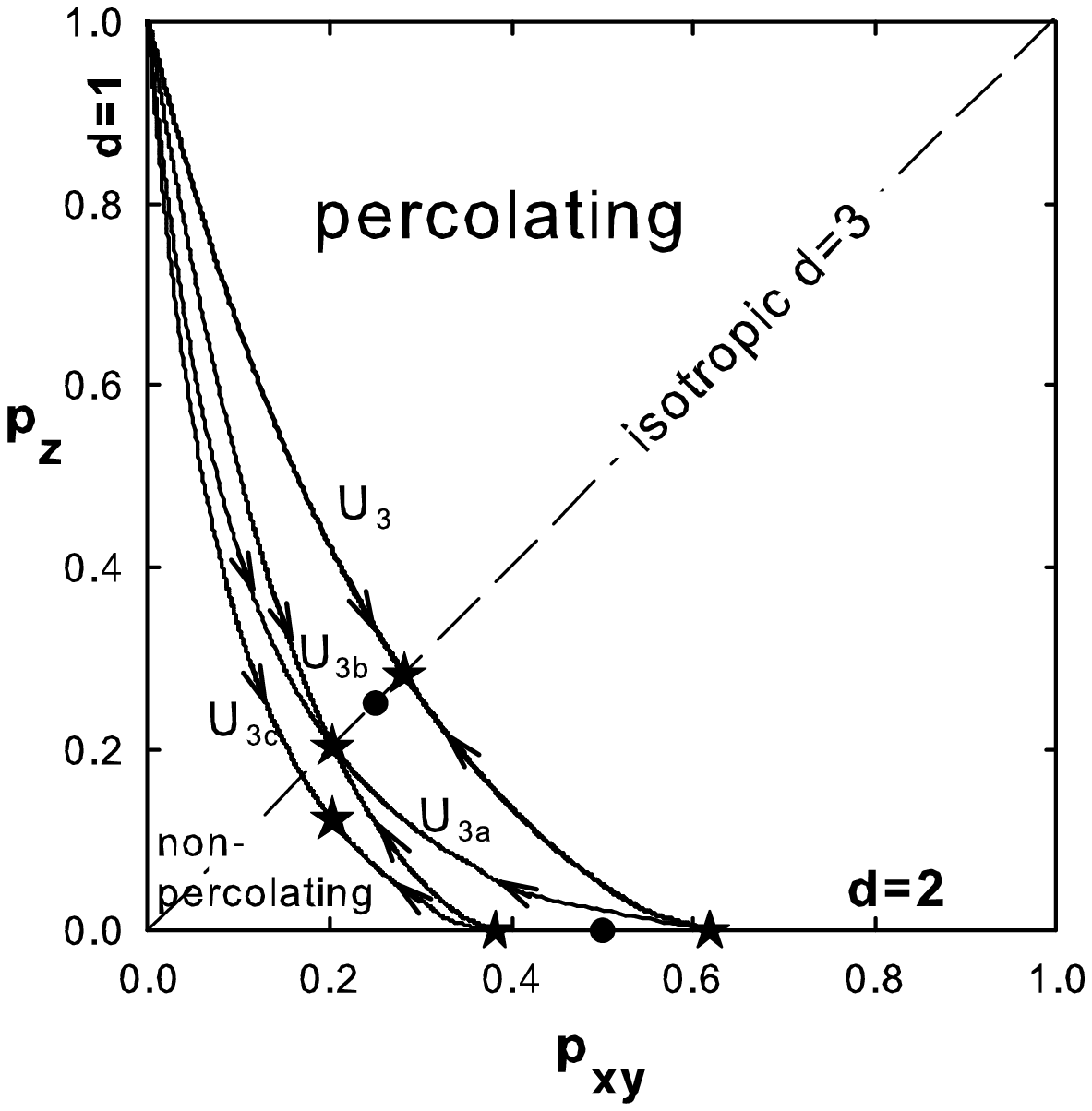}

\caption{Phase boundaries for percolation on the $d=3$ uniaxially
anisotropic hierarchical lattices $U_3, U_{3a}, U_{3b}$, and
$U_{3c}$.  The onsets of percolation for the square and cubic
lattices \cite{Essam} are shown by the black circles.  For each
model, the percolation onset at $d=1$ (at $p_z =1$) and $d=2$ cross
over to $d=3$ percolation onset, which is thus universal for all
$d=3$ anisotropic and the $d=3$ isotropic cases.}
\end{figure}

This research was supported by the Scientific and Technical Research
Council of Turkey (T\"UBITAK) and by the Academy of Sciences of
Turkey.

\end{document}